# IRHA: An Intelligent RSSI based Home automation System

1.Samsil Arefin Mozumder[1] and A S M Sharifuzzaman Sagar[2]

[1] East Delta University, Chattogram, Bangladesh
[2] Sejong University, Seoul, South Korea
`samsilarefin313@gmail.com`
`sharifsagar80@sju.ac.kr`

**Abstract.** Human existence is getting more sophisticated and better in many areas due to remarkable advances in the fields of automation. Automated systems are favored over manual ones in the current environment. Home Automation is becoming more popular in this scenario, as people are drawn to the concept of a home environment that can automatically satisfy users' requirements. The key challenges in an intelligent home are intelligent decision making, location-aware service, and compatibility for all users of different ages and physical conditions. Existing solutions address just one or two of these challenges, but smart home automation that is robust, intelligent, location-aware, and predictive is needed to satisfy the user's demand. This paper presents a location-aware intelligent RSSI based home automation system (IRHA) that uses Wi-Fi signals to detect the user's location and control the appliances automatically. The fingerprinting method is used to map the Wi-Fi signals for different rooms, and the machine learning method, such as Decision Tree, is used to classify the signals for different rooms. The machine learning models are then implemented in the ESP32 microcontroller board to classify the rooms based on the real-time Wi-Fi signal, and then the result is sent to the main control board through the ESP32 MAC communication protocol to control the appliances automatically. The proposed method has achieved 97% accuracy in classifying the users' location.

## 1    Introduction

The home automation system enables users to operate various types of devices and allows managing home appliances to be more straightforward and saves energy. Automation systems for the home and building are becoming popular [1]. On the other side, they improve comfort, mainly when everyone is preoccupied with work. "Home automation systems" placed in homes boost comfort in addition to providing centralized management of airflow, heating, air conditioning, and lighting [2].
In the last ten years, academics have introduced a slew of home automation systems. Wireless home automation systems have made use of a variety of technologies, each with its own set of benefits and drawbacks. As an example, Bluetooth-based automation can be quickly and easily installed and de-installed, but it is limited to short distances. GSM and ZigBee are additional extensively used wireless technologies. A local service provider's phone plan is required to use GSM for long-distance communication. For battery-powered devices in wireless real-time applications, Zigbee [6–11] is a wireless mesh network protocol developed for low-cost, low-energy usage. Data rates, transmission, and network dependability are all constrained, and upkeep is prohibitively expensive. Wi-Fi technology is used in [9,11,12-14]. Wi-Fi technology has advantages over ZigBee and Z-Wave in terms of cost, ease of use, and connection. With Wi-Fi equipped smart gadgets, the cost is usually low. Do-it-yourself Wi-Fi equipment is also easier to come by, resulting in an affordable option. Second, because Wi-Fi is now required and installed in the majority of homes, buying appliances that are Wi-Fi enabled is easier. As Wi-Fi is noted for its simplicity, a user only has to connect a limited number of devices for a home automation system. As a result, it seems that home automation systems using Wi-Fi are a better fit.

On the other hand, the elderly constitutes a significant and rising part of the global population. Statistics reveal that the proportion of persons aged 65 and over is steadily increasing due to a variety of factors, including decreased birth rates and women's fertility. The percentage of the population aged 65 and above in the United States climbed from 12.4 percent in 2000 to 13.3 percent in 2011, and it is anticipated to rise to 21 percent by 2040 [15]. According to a United Nations study [16], life expectancy was 65 years in 1950, 78 years in 2010, and is expected to climb to 83 in 2045. Moreover, it was reported that 35% of people aged 65 and above had a handicap [15]. Some of them need help to achieve critical personal requirements. Adopting home automation systems with automatic essential control appliances can minimize the cost of in-home personal assistance.

To ensure that home automation systems are suitable for all users, critical requirements such as cost-effective, location-aware service, automatic control appliances, wireless connectivity must be fulfilled. This paper presents practical solutions to fulfill the requirements by undertaking the following methods.

1. The proposed system is cost-effective because the IP-based light and bulb were not used as they were controlled using a relay module.
2. The proposed system uses a Wi-Fi signal to detect the user's location through a machine learning approach.
3. The proposed system can automatically control appliances without any human interference.
4. The proposed system has Wi-Fi connectivity to send the control signal to the main control unit to control the appliances.
5. The proposed approach delivers a user-friendly prototype of a home automation system.

The rest of the paper is divided as follows: related works relevant to home automation system is discussed in section 2, section 3 describes the method used in the proposed system, section 4 presents the proposed system implementation in a home, the results of the proposed system implementation is discussed in section 5, and finally conclusion is drawn in section 6.

## 2 Related work:

Table 1 shows the research gaps highlighted in earlier studies in the domains of home automation systems. The research gap in current studies in the domains of smart homes, smart buildings, and smart surroundings is shown in Table 1. Location aware service needs precise location information to enable services. Previous methods are based on the instruction provided by the users to provide home automation services. However, elderly, handicapped and Childs may find difficult in controlling the appliances through app or manually.

*Table 1 Previous studies and research gap in the field of home automation system*

| Home Automation System | Communication | Controller | Application |
|---|---|---|---|
| 3 | Bluetooth | PIC | Control indoor appliances |
| 4 | Bluetooth | Arduino | manage appliances both indoors and outdoors at a short distance |
| 5 | Bluetooth, GSM | PIC | operate both indoor and outdoor devices |
| 6 | ZigBee, Ethernet | Arduino | operate indoor devices |
| 7 | X10, Serial, EIB, ZigBee, Bluetooth | ARM MCU | Indoor home automation service |
| 8 | Wi-Fi, ZigBee | Raspberry PI | Controlling humidity, temperature |
| 9 | ZigBee | Laptop | Control of home appliances is being discussed, but no implementation has not been taken. |
| 10 | Wi-Fi | Raspberry PI board | A/C system implementation |

## 3   Materials and Methods:

### 3.1   System Architecture and Design

The proposed home automation system can be divided into data acquisition, prediction based on the data, and control appliances automatically. Fig. 1 shows the basic working architecture of the Intelligent RSSI based home automation system (IRHA) system. The proposed system is consisting of a data acquisition unit and control unit. The data acquisition unit is designed as a wristband that the user can wear. The ESP32 development board is used in the data acquisition unit to receive the Wi-Fi signal (RSSI) from the Wi-Fi routers, and the RSSI signals are fed into the machine learning classifier to detect the user's location. The decision tree machine learning model trains and detects the user's location based on the RSSI signals. The control data is sent to the main control unit through the ESP-NOW communication protocol to control the appliances. The main control unit is consisting of an ESP32 development board, relay, and power connection. The control unit receives the predicted location data and controls the appliances based on the location data. The relay modules are connected with appliances such as bulb, fan and when the user is in the specific room relay turn on the appliances.

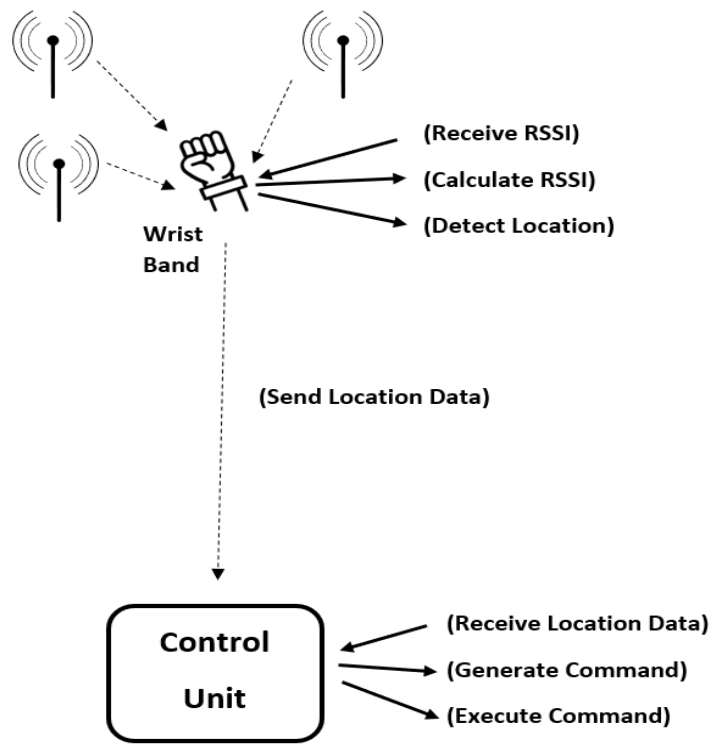

Figure 1 The basic architecture of the IRHA system.

## 3.2 Wi-Fi fingerprinting:

Wi-Fi Fingerprinting develops a probability distribution of RSSI values for a particular place and builds a radio map of a particular region based on RSSI data from multiple gateways [19]. The closest match is then identified, and a projected location is generated by comparing current RSSI readings to the fingerprint. In principle, this technique's implementation may be separated into two different phases:
1) Offline phase: fingerprints of received signal intensities are gathered from several accessible gateways for distinct reference locations whose locations have previously been calculated and then saved in a system.
2) Online phase: A pattern-matching technique is used to compare a user's fingerprint to those in the database while the user is logged in online. The user's location is then determined to be the one that corresponds to the source database's nearest fingerprint.

## 3.3 Decision Tree classification

A decision tree is a prediction model that is often used to illustrate a classification strategy. Classification trees are a popular exploration tool in many industries, including economics, trade, medicine, and engineering [20, 21, 22, 23]. The decision tree is often used in data analysis because of its efficiency and clarity. Decision trees are visually depicted as hierarchical, making them simpler to understand than alternative strategies. This structure primarily consists of a root node and a series of forks (conditional) that result in additional nodes until we approach the leaf node, which holds the route's ultimate conclusion. Due to the simplicity of its depiction, the decision tree is a self-explanatory structure. Each internal node evaluates one attribute, whereas each branch represents the property's value. Finally, each lead gives a classification to the scenario.

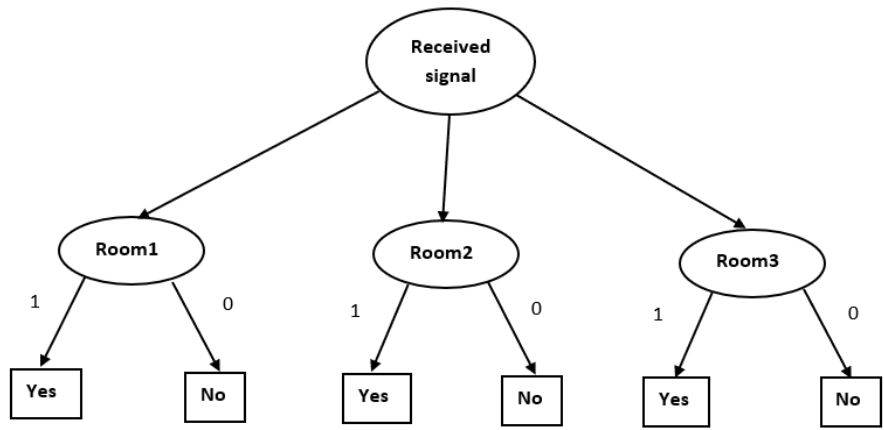

Figure 2 Decision tree example for IRHA system

Fig. 2 shows an example for a basic decision tree "user location" classification model used in our proposed method. It simply decides whether the user is in a particular room based on the different room's RSSI signature. RSSI values of different rooms are collected from the nearby Wi-Fi access point using the ESP32 Wi-Fi module and then saved in the database. We have used the scikit-learn machine learning python library to train our dataset for location detection. The datasets were split into the 70 30 training and testing ratios, respectively. The trained model is then ported to the microcontroller using the elequentarduino library.

## 3.4 ESP-now protocol:

ESP-NOW is an Espressif-defined protocol for connectionless Wi-Fi connectivity. Data packet is contained in a vendor-specific action frame and sent through one Wi-Fi device to the other without the need for connectivity in ESP-NOW. The CTR with CBC-MAC Protocol (CCMP) is used to secure the action frame in the ESP-now approaches-NOW uses minimal CPU and flash memory, can connect to Wi-Fi and Bluetooth LE, and is compatible with different ESP families. ESP-NOW has a versatile data transmission method that includes unicast, broadcast, and one-to-many and many-to-many device connections. ESP-NOW can also be used as a stand-alone supplemental module to assist network configuration, diagnostics, and firmware updates.
ESP-NOW defines two roles, initiator and responder, based on the data flow. The same device may perform two functions concurrently. In an IoT infrastructure, switches, sensors, LCD panels, and other real-time applications often serve as the initiator, while lights, sockets, and other real-time applications serve as the responder, according to the industry standard. ESP-NOW protocol provides 1Mbps of bit rate by default. We have used the ESP-NOW protocol for our proposed system as it is lightweight, less storage-consuming.

## 4  Implementation

### 4.1  Hardware Components:

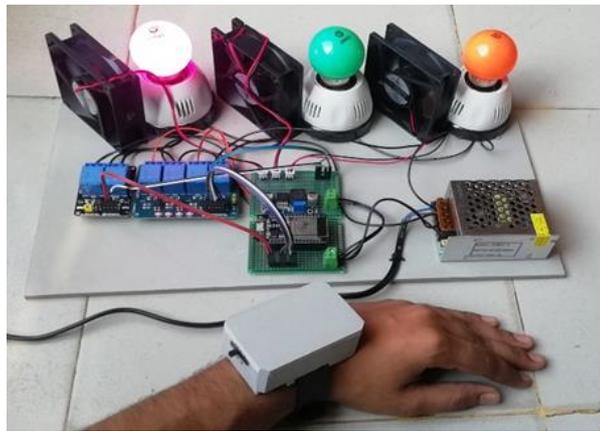

*Figure 3 The hardware components of the proposed home automation system*

The hardware components mainly consist of two parts: Wrist Band & Control System. Fig.3 shows the hardware components used in the proposed system. Both parts have ESP32 and power sources in common. In the case of the wristband, a tiny battery has been used to power it up. The Control system uses SMPS to activate the whole unit. It has 3 Lights and 3 Fans, and relay modules control those. ESP32 provides the control signal for relay modules. A detailed description of the used hardware is presented below:

ESP32 development board: ESP32 development board, the successor of the ESP8266 kit. The ESP32 has several additional features compared to its predecessor. It has a dual-core processor and wireless connectivity through Wi-Fi and Bluetooth.

Relay Module: Microcontroller output signals cannot run real-world loads, so a relay module has been introduced to control the appliances. Initially, output signals from the microcontroller unit are sent to a photocoupler to provide base voltage to the driver transistor of the relay. 6 channel relays have been used in our proposed system.

## 4.2 Control system method:

**Algorithm 1:** Intelligent RSSI based home automation system control procedure

**Input** : RSSI
**Output:** Light control, Fan control

1 $var \leftarrow RSSI$
2 $var \rightarrow classifier$
3 $classifier \rightarrow user's location$
4 $control - var \leftarrow user's location$
5 $control - var \rightarrow control - unit$
6 **if** $control - var = 1$ **then**
7 $\quad$ Turn on light and fan of room 1
8 **else if** $control - var = 2$ **then**
9 $\quad$ Turn on light and fan of room 2
10 **else if** $control - val = 3$ **then**
11 $\quad$ Turn on light and fan of room 3
12 **else**
13 $\quad$ Turn off all actuators

Algorithm 1 shows the control system of the proposed home automation system. The wristband collects the RSSI data from the nearby Wi-Fi access points as input and saves the variable's data. The acquired RSSI value is then fed into the classifier to determine the user's location. When the system detects the y=user's location, it saves a specific control signal in the control variable based on the room. The ESP-NOW communication protocol sends the control signal to the main control unit. The control unit compares the control signal with the prespecified signal to control the relevant room to enable the home automation system.

## 5 Results and Discussion:

### 5.1 Decision Tree classification

Decision tree classification has been used in the proposed system to classify the RSSI value to determine the user's location. Data acquisition and pattern recognition are among the most important factors for a Wi-Fi-based fingerprinting technique. Data acquisition was made by taking the RSSI value for every room using the ESP32 development board. The acquired data are then used to train the classifier models to determine the best models. We have trained our dataset with available and common classifiers such as SVM, Decision Tree, Random Forest, and Naïve Bayes. We have found that Decision Tree outperforms other algorithms in terms of accuracy, which is 97%. Therefore, we used the Decision Tree classifier for our proposed home automation system. Figure 4 shows a trained Decision Tree confusion matrix, which indicates the percent of correctly predicted and incorrectly predicted classes. The confusion matrix depicts occurrences of the true label horizontally and predicted label instances vertically. There was a total of 19 expected "room1" classes, while the proposed model predicted all instances correctly for room1". As for room 2, the proposed model predicts 2 incorrect labels, while the proposed model predicts all classes correctly for room3.

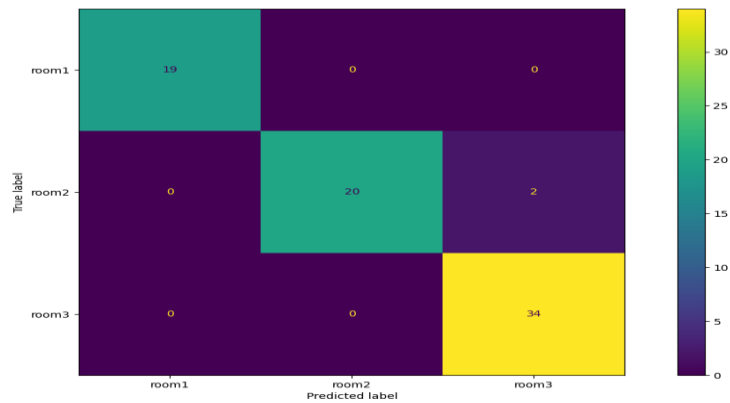

Figure 4 Confusion matrix of the proposed model

We also present the Decision Tree classifiers ROC curve in figure 5 for every class. The true positive rate of the proposed model is displayed on the y-axis in the ROC curve, while the false-positive rate is represented on the x-axis. Micro average and macro average ROC curves are calculated for the proposed model. We have also calculated ROC and AUC for every class for our proposed model.

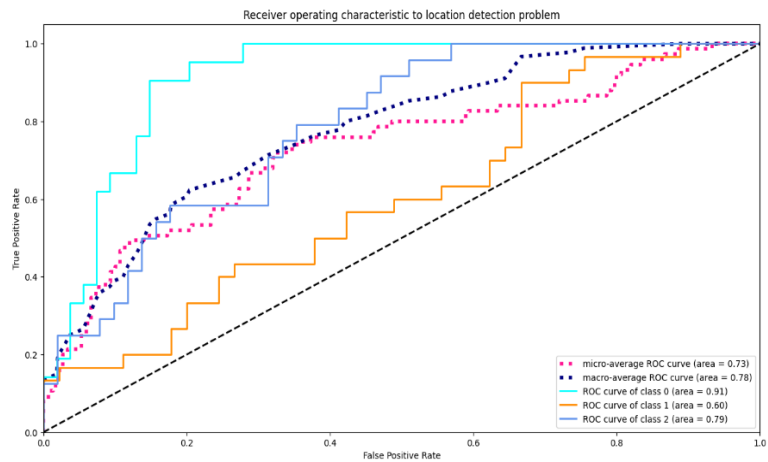

Figure 5 ROC curve for all classes of the proposed model

## 6 Conclusion:

The whole design plan and working technique of a home automation system is described in detail in this research study. The purpose of the paper is to seek ideas about how to improve it and make it more user-friendly, particularly for the elderly and disabled. This suggested system has two mechanisms: data acquisition and control. The system collects Wi-Fi RSSI values to anticipate the user's position on the data acquisition side, and on the control side, the system automatically controls appliances. A machine learning method called a Decision tree classifier is used to predict the state of the user's location. The proposed method also uses the ESP-NOW protocol to enable a secure connection between the data acquisition and control units. It offers users simplicity, flexibility, and dependability, as well as a low-cost solution.